\documentclass[aps,prl,preprint,superscriptaddress]{revtex4-1}

\usepackage{graphicx}
\usepackage{dcolumn}
\usepackage{bm}
\usepackage{hyperref}

\begin{document}


\title{Limits for superfocusing with finite evanescent wave amplification}

\author{Reuven Gordon}
 \email{rgordon@uvic.ca}
 \affiliation{Department of Electrical and Computer Engineering, University of Victoria}

\begin{abstract}
Perfect lensing using negative refractive index materials and radiationless electromagnetic interference both provide extreme subwavelength focusing by ``amplifying" evanescent wave components that are usually lost. This paper provides a relation between the achievable focus spot size, the amplification available and the focal length. This may be considered as a revised version of Abbe's diffraction limit for focusing systems that have evanescent wave amplification. It is useful in comparing the amplification
achieved in various subwavelength focusing implementations, as well as determining when it is better to use existing near-field techniques, such as simple diffraction from an aperture or slit, than to attempt complicated superfocusing.
\end{abstract}


\maketitle

\noindent In 1873, Abbe stated that ``No particles can be resolved ... when they are situated so close together that not even the first of a series of diffraction pencils produced by them can enter the objective simultaneously with the undiffracted rays."~\cite{abbe1873} (as translated in Ref.~\cite{fripp1874}). This may be formulated as Abbe's resolution limit relating the resolution of a microscope, $d$ to its numerical aperture, NA, and the optical wavelength, $\lambda$ as:
\begin{equation}
d = \frac{\lambda}{2 \mathrm{NA}}
\label{abbe}
\end{equation}
as recorded by Helmholtz in 1876~\cite{helmholtz1876}. To within a factor of the order of unity, which is a matter of lineshape and definition of linewidth, this relation also dictates the smallest focus spot that can be produced by a microscope.

The connection of Abbe's resolution limit to the ``uncertainty principle" is common~\cite{novotny2006,lipson2011}; however, care should be taken here because a conventional imaging system is strictly band-limited in Fourier space and so the standard deviation which is commonly used as the ``uncertainty" is not defined in real space. By strictly band-limited, I mean that wave-vector components greater than $k_0 = 2 \pi/\lambda$ are zero. As a possible substitute, the mean absolute deviation can be used, or the full-width at half-maximum (FWHM).

Many schemes exist to surpass Abbe's resolution limit within the context of conventional imaging~\cite{lipson2011}. These schemes can make use of Fourier information from a finite spatial extent that leaks into the allowed band of the imaging system -- for example, with stochastic optical reconstruction microscopy, the position of a point source can be determined accurately and several such point source labels can be pieced together to reconstruct an image~\cite{storm}. It is possible to have super-oscillations in a band-limited system, and this allows for creating a local peak with a small spot-size~\cite{lipson2011}; however, this comes at the expense of having large side lobes. Nonlinear processes, such as multi-photon imaging~\cite{nonlinear} and stimulated emission depletion~\cite{sted}, which uses the nonlinearity in fluorescence depletion, also allow for super resolution.

It is possible to surpass the band limitation of conventional imaging systems by recovering wave-vector components $k > k_0$. One way to do this is to operate in the near-field, as with a near-field scanning optical microscope~\cite{nsom1,nsom2}, where these evanescent components can be captured before they decay away. This requires getting really close to the object to be imaged. To allow for a subwavelength focus spot at a larger distance (for example, to be less invasive), schemes have been proposed which amplify the evanescent components. One scheme is the so-called ``perfect lens" using negative index materials~\cite{pendry2000}. Another is boost the evanescent components with respect radiating modes prior to allowing for diffraction in ``radiationless electromagnetic interference"~\cite{merlin2007}. This paper derives a relation between the achievable focus spot size, the ``amplification" available and the focal length.

We consider the 2D case for simplicity. Light propagation along the $z$ direction may be decomposed into its plane wave components, each with transverse wave-vector $k_x$. The propagation of each of these transverse wave-vector components is expressed as $\exp(i k_z(k_x) z)$, where $i = \sqrt{-1}$ and $k_z = \sqrt{k_0^2 - k_x^2}$. For $k_x > k_0$, $k_z$ is imaginary and these components decay away exponentially. Therefore, the natural transverse wave-vector bandwidth for light propagation is $k_0$. If we wish to extend this bandwidth to $k_x'$ by amplification, the amplification required is:
\begin{equation}
G(k_x') = \exp(-i k_z(k_x') z)
\end{equation}
which we may rearrange to write as:
\begin{equation}
k_x' = \sqrt{\left(\frac{\ln G(k_x')}{z}\right)^2 + k_0^2}.
\label{kx}
\end{equation}
Through the Fourier transform, there is a natural relation between the transverse bandwidth and the transverse resolution (i.e., spot-size), $d$, that can be obtained. The details of this relation will depend on lineshape, and we will consider the case of a Gaussian lineshape further below, but for now we will simply state the relation as:
\begin{equation}
k_x' d = \pi
\label{ur1}
\end{equation}
which combined with the previous equation gives:
\begin{equation}
d = \frac{\pi}{\sqrt{\left(\frac{\ln G(k_x')}{z}\right)^2 + k_0^2}}.
\label{re1}
\end{equation}
This may be viewed as a relation for the achievable spot-size in the presence of amplification of the evanescent components. Now we see that the choice of $\pi$ in Eq.~\ref{ur1} is motivated by the limit $z$ sufficiently large, or the amplification is sufficiently small, such that $k_0$ dominates the denominator of Eq.~\ref{re1} and so:
\begin{equation}
d \rightarrow \lambda/2
\end{equation}
which is again Abbe's resolution limit of Eq.~\ref{abbe} for NA unity. Therefore, a nice feature of composing the bound on focus spot this way is that it approaches Abbe's limit in the long distance limit, which is not the case for the Gaussian lineshape derivation that will be given below because it is not band-limited.

Of course, most subwavelength focusing with operate at short distances $z \ll \lambda$, so we consider the opposite limit with sufficient amplification where
\begin{equation}
g = \ln G(k_x') \rightarrow \pi z/d.
\label{lim1}
\end{equation}
This equation basically tells us that the amount of amplification scales as $z/d$. In other words, when comparing superfocusing schemes, such as ``perfect lensing" or ``radiationless electromagnetic interference", a suitable metric is $z/d$. This is a rather intuitive result that tells you that it is more difficult to achieve a narrower focus spot for a longer distance (i.e., it requires more amplification), and more specifically, these two quantities are linearly related.

Next we consider specifically the Gaussian lineshape, which minimizes the uncertainty relation but is not applicable to band-limited systems. The amplification required for this case is:
\begin{equation}
G(k_x) = \exp\left(-i k_z(k_x) z - \frac{k_x^2}{2 \sigma_k^2}\right)
\end{equation}
where $\sigma_k$ is the standard deviation of the transverse Fourier spectrum (with respect to the field, not the intensity).
From this we find the maximum amplification required is:
\begin{equation}
g_\mathrm{max} = \ln G_\mathrm{max} = \frac{z^2 \sigma_k^2}{2} + \frac{k_0^2}{2 \sigma_k^2}
\end{equation}
The standard deviation with respect to the intensity has the relation:
\begin{equation}
\sigma_k \sigma_x = 1
\end{equation}
where $\sigma_x$ is the standard deviation of the intensity in real space, which we write as $d$ to be consistent with the notation above. This is the minimal form of the uncertainty relation which is found for Gaussian distributions.
Therefore,
\begin{equation}
g_\mathrm{max} = \frac{z^2}{2 d^2} + \frac{k_0^2 d^2}{2 }
\end{equation}
which again gives the result for small enough $d$ (ignoring the second term) that $z/d$ is the critical metric for amplification; however, now the form of the amplification scales as the square of this ratio.

It is interesting to consider when it is better to use amplification of the evanescent waves to achieve a subwavelength spot, as opposed to, for example, a simple single slit in a metal film. An $y$-oriented infinitessimal slit in a metal film has TM transmission for which the electric field component $E_x$ scales as $1/(x^2 + z^2)$. The FWHM in the transverse direction $x$ at a distance $z$ is given by $2 z$, which corresponds to:
\begin{equation}
g = \pi/2
\label{slit}
\end{equation}
from Eq.~\ref{lim1}. This will be the baseline for comparison.

Next we consider the focusing of TM waves achieved by a thin silver film in the electrostatic regime (i.e., near-field) as described in Ref.~\cite{pendry2000}. This case has been referred to as the poor man's perfect lens. Using the approach of that work it may be shown that the transverse wave-vector components traveling through free-space of thickness $z$, a silver slab of thickness $2z$ and then free-space again of thickness $z$ would be subject to the propagator:
\begin{equation}
G(k_x) \exp\left(i k_z z\right) = \frac{4 \epsilon}{(\epsilon + 1)^2 \exp\left(-2 i k_z z\right) - (\epsilon - 1)^2}
\end{equation}
For $\epsilon = -1 + i \epsilon''$, where $\epsilon''$ is the imaginary part of the relative permittivity, assuming $\epsilon'' \ll 1$ and considering the quasistatic regime where $k_z = i k_x$, this gives:
\begin{equation}
G(k_x) \exp\left(i k_z(k_x) z\right)\approx \frac{1}{(\epsilon''/2)^2 \exp\left(2 k_x z\right) + 1}
\end{equation}
From this, the bandwidth is approximately given by $k_x' = \ln(2/\epsilon'')/z$ (when the denominator equals 2), or considering Eq.~\ref{kx}:
\begin{equation}
g = \ln(2/\epsilon'').
\end{equation}
To surpass the single slit case then, requires $\epsilon'' < 2 \exp(-\pi/2) = 0.41$. Typical values for silver give  $\epsilon'' = 0.3$ when the real part is -1~\cite{jc}, which is slightly better than this approximate limit. Fig. 1 shows a comparison of the diffraction from an infinitessimal slit with and without focusing by a 40 nm metal slab for different values of $\epsilon''$, for the same working distance. While the lineshape is clearly different between the diffraction and the focusing cases, and the approximate analysis above does not account for these differences, it is clear that $\epsilon'' \sim 0.1$ is the order required for the losses to benefit from focusing with the same working distance.

\begin{figure}[htb]
\centerline{
\includegraphics[width=8.3cm]{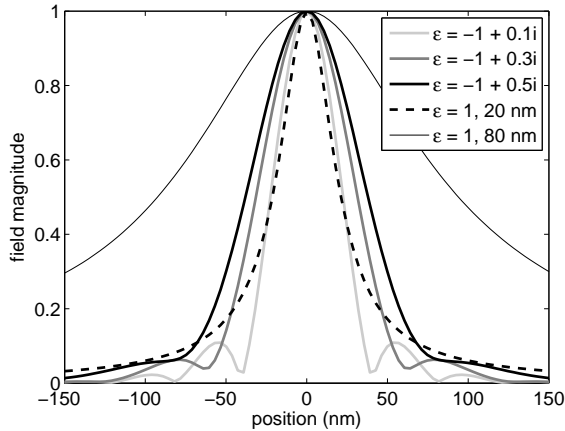}}
 \caption{Comparison of a subwavelength focus achieved using a 40 nm metal slab of varying permittivity (particularly $\epsilon''$), excited with a magnetic line source 20~nm away and imaging at the focal plane 20~nm away from the other side of the slab. For comparison, the free-space ($\epsilon = 1$) diffraction of the line source at a distance of 20~nm and 80~nm are shown.}
\end{figure}

Next we consider demonstrations of radiationless electromagnetic interference to produce a subwavelength focus spot in the microwave regime. For near field plates at approximately 1~GHz frequency, focusing to a intensity FWHM of $\lambda/20$ at a distance of $\lambda/15$ was achieved~\cite{grbic2008}. Assuming a Gaussian profile, this corresponds to a electric field FWHM of $\lambda/14$. Another work using 3 slots in a metal plate produced a subwavelength focus of $\lambda/3.7$ at a distance of $\lambda/4$~\cite{markley}. Considering the metric of $z/d$, these two experimental works produce nearly identical results in terms of the evanescent wave amplification achieved: $g = 0.93 \pi$. This similarity occurs even though the latter operates for a much larger focal length and has a much larger spot-size. Perhaps the similarity is not surprising since they are using the same region of the electromagnetic spectrum and they use the same concept of radiationless intereference (although the latter work is framed in terms of general interference). This idea of evanescent amplification is useful as a metric in comparing focusing at different distances.

In summary, this work suggests a new version of Abbe's limit for superfocusing in the presence of evanescent wave amplification. This limit allows us to compare different methods of superfocusing by comparing the focal length divided by the spot-size as the critical parameter, which can be linearly or quadratically related to the amplification, depending on chosen formulation. It also allows us to consider when superfocusing would be superior to allowing for diffraction from, for example, a simple slit.

R. G. acknowledges financial support from the Natural Sciences and Engineering Research Council of Canada Discovery Grant. R. G. thanks Dr. P. J. Schuck for reviewing an early version of this manuscript and providing useful suggestions for improvement.


\end{document}